\documentclass[sigconf]{acmart}

\AtBeginDocument{%
  }

\copyrightyear{2025}
\acmYear{2025}
\setcopyright{cc}
\setcctype[4.0]{by}
\acmConference[CIKM '25]{Proceedings of the 34th ACM International
Conference on Information and Knowledge Management}{November 10--14,2025}{Seoul, Republic of Korea}
\acmBooktitle{Proceedings of the 34th ACM International Conference on
Information and Knowledge Management (CIKM '25), November 10--14, 2025,
Seoul, Republic of Korea}
\acmDOI{10.1145/3746252.3760908}
\acmISBN{979-8-4007-2040-6/2025/11}

\newcommand{\IGNORE}[1]{}

\def\tabref#1{Table~\ref{#1}}

\usepackage{booktabs}
\usepackage{multicol}
\usepackage{multirow}
\usepackage{adjustbox}
\usepackage{subfig}
\usepackage{balance}

\begin{document}

\title{Towards Understanding Bias in Synthetic Data for Evaluation}


\author{Hossein A.~Rahmani}
\orcid{0000-0002-2779-4942} 
\affiliation{%
        \institution{University College London \\ The Alan Turing Institute}
        \city{London}
        \country{UK}
}
\email{hossein.rahmani.22@ucl.ac.uk}

\author{Varsha Ramineni}
\orcid{0000-0002-0330-3184} 
\affiliation{%
        \institution{University College London}
        \city{London}
        \country{UK}
}
\email{varsha.ramineni.23@ucl.ac.uk}

\author{Emine Yilmaz}
\orcid{0000-0003-4734-4532} 
\affiliation{%
        \institution{University College London}
        \city{London}
        \country{UK}
}
\email{emine.yilmaz@ucl.ac.uk}

\author{Nick Craswell}
\orcid{0000-0002-9351-8137} 
\affiliation{%
        \institution{Microsoft}
        \city{Seattle}
        \country{US}
}
\email{nickcr@microsoft.com}

\author{Bhaskar Mitra}
\orcid{0000-0002-5270-5550} 
\affiliation{%
        \institution{Microsoft}
        \city{Montréal}
        \country{Canada}
}
\email{bmitra@microsoft.com}

\renewcommand{\shortauthors}{Hossein A.~Rahmani, Varsha Ramineni, Emine Yilmaz, Nick Craswell, and Bhaskar Mitra}

\begin{abstract}
Test collections are crucial for evaluating Information Retrieval (IR) systems. Creating a diverse set of user queries for these collections can be challenging, and obtaining relevance judgments, which indicate how well retrieved documents match a query, is often costly and resource-intensive. Recently, generating synthetic datasets using Large Language Models (LLMs) has gained attention in various applications. While previous work has used LLMs to generate synthetic queries or documents to improve ranking models, using LLMs to create synthetic test collections is still relatively unexplored. Previous work~\cite{rahmani2024synthetic} showed that synthetic test collections have the potential to be used for system evaluation, however, more analysis is needed to validate this claim. In this paper, we thoroughly investigate the reliability of synthetic test collections constructed using LLMs, where LLMs are used to generate synthetic queries, labels, or both. In particular, we examine the potential biases that might occur when such test collections are used for evaluation. We first empirically show the presence of such bias in evaluation results and analyse the effects it might have on system evaluation. We further validate the presence of such bias using a linear mixed-effects model. Our analysis shows that while the effect of bias present in evaluation results obtained using synthetic test collections could be significant, for e.g.~computing absolute system performance, its effect may not be as significant in comparing relative system performance. Codes and data are available at: \url{https://github.com/rahmanidashti/BiasSyntheticData}
\end{abstract}




\maketitle

\section{Introduction}
\label{sec:introduction}
The development of test collections is foundational for evaluating information retrieval systems. Traditionally, constructing these collections has relied on the Cranfield paradigm \cite{cleverdon1967cranfield,harman2011information,sanderson2010test}, which requires substantial human effort to create queries and associated relevance judgments, making the process both costly and time-consuming \cite{aslam2006statistical,sanderson2010test}. With the rise of Large Language Models (LLMs), there has been growing interest in leveraging these models to automate and reduce the cost of test collection construction \cite{rahmani2025report,rahmani2025judgeblender}. LLMs have demonstrated impressive capabilities across a wide range of tasks \cite{gao2024llm,li2024leveraginglargelanguagemodels,wang2023chatgpt}, from generating synthetic training data \cite{bonifacio2022inpars,askari2023expand} to providing relevance judgments for search and retrieval systems \cite{faggioli2023perspectives,farzi2024pencils,thomas2024large}, suggesting that they could effectively streamline the test collection creation process.

Recent advances have shown that LLMs can be employed to generate synthetic test collection, including queries and relevance judgments \cite{rahmani2024synthetic,rahmani2024syndl,rahmani2024llmjudge,rahmani2025llm4eval}. This synthetic approach offers a significant reduction in cost and time compared to traditional methods, and it has the potential to enhance the scalability of test collection construction. However, the reliability and accuracy of LLM-generated data for system evaluation purposes remain important areas of concern. LLMs are inherently influenced by the data they are trained on \cite{liu2023llms,wang2023large}, which can introduce biases into synthetic evaluations. This presents challenges in using LLM-generated test collections as a reliable substitute for human-annotated benchmarks, particularly when biases could impact the fairness and effectiveness of retrieval systems.

\begin{figure}
    \centering
    \subfloat[Query Length\label{fig:query_length}]
    {
        {
            \includegraphics[width=0.45\linewidth]{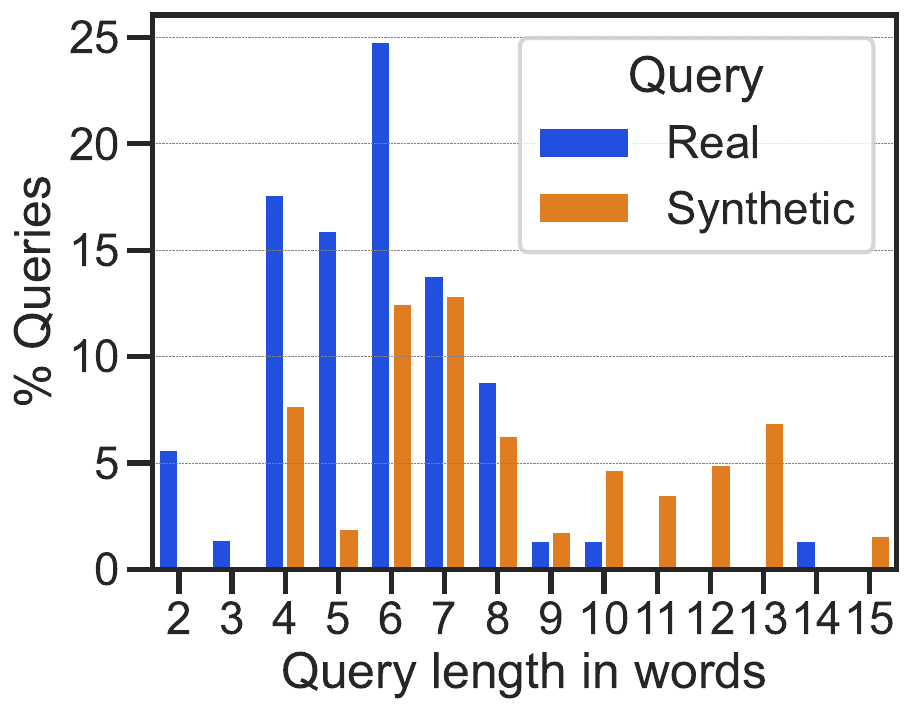}
        }
    }%
    \subfloat[Label Differences\label{fig:label_diff}]
    {
        {
            \includegraphics[width=0.55\linewidth]{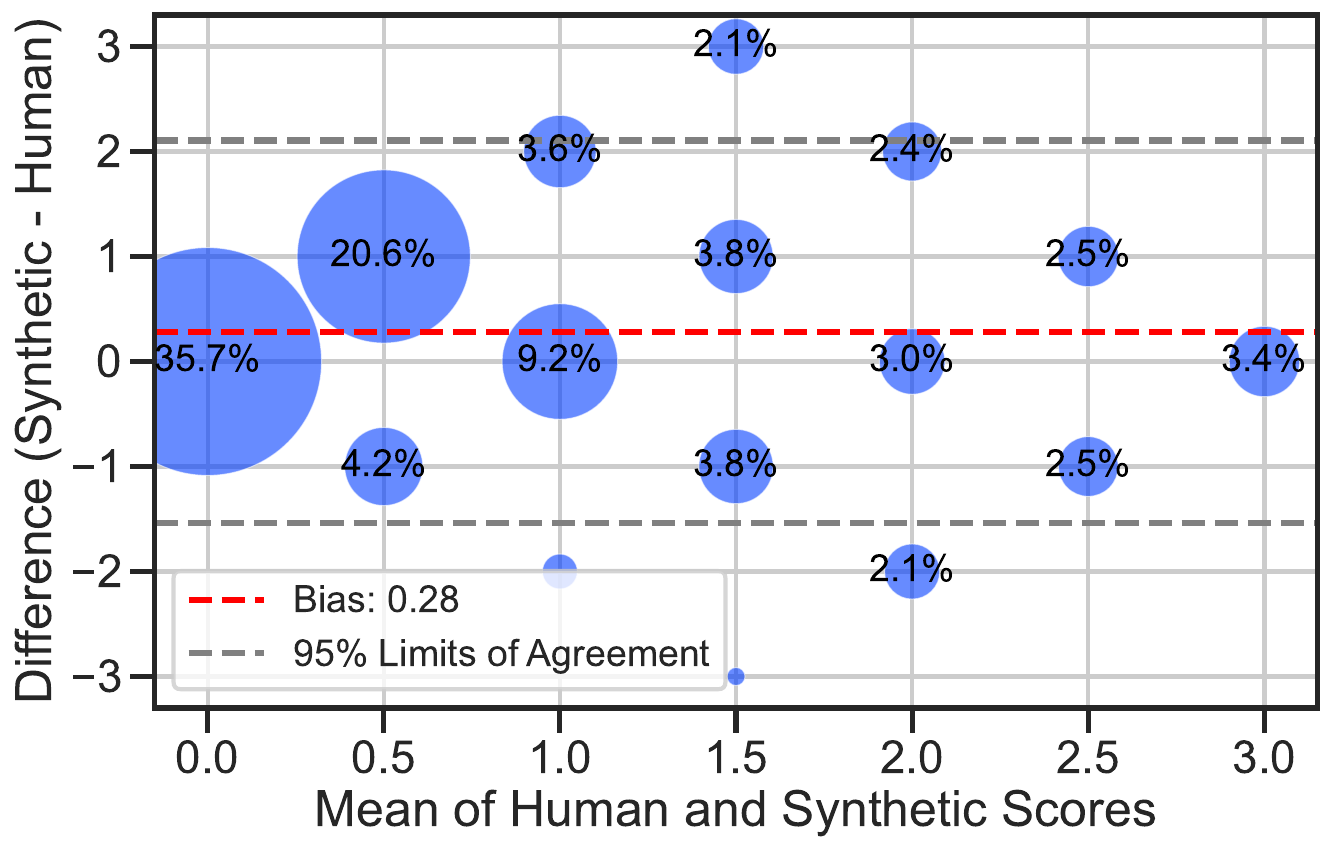}
        }
    }
    \caption{(a) The percentage of queries based on the number of words in the queries. Real queries are shorter than synthetic queries. (b) Bland-Altman plot to visualize the comparison between LLM and human expert judgments.}
    \label{fig:query-length-precentage}
\end{figure}

While some studies have investigated the general reliability of LLMs in providing synthetic relevance judgments \cite{dietz2025llm,balog2025rankers}, there is no study on the analysis of the biases introduced by synthetic test collections. Understanding such biases is crucial, especially in scenarios where synthetic collections are used to evaluate retrieval systems that may later be deployed in high-stakes or user-facing applications. Bias in synthetic data can lead to skewed evaluations, which may, in turn, affect the perceived quality and trustworthiness of information retrieval systems. In this paper, we conduct an in-depth bias analysis of synthetic test collections constructed using LLMs. Specifically, we aim to identify and characterize the types of biases that arise when LLMs are used to generate both queries and relevance judgments, and how these biases impact system evaluation. We examine differences in characteristics of human and synthetic queries and judgments, and consistency in synthetic judgments, thereby providing insights into the broader implications of employing LLMs for constructing evaluation test collections. Our findings contribute to a better understanding of the strengths and limitations of synthetic test collections. Our results from the linear mixed-effect model indicate that LLM-based systems receive higher scores when evaluated on synthetic judgment.
\section{Synthetic Test Collection Dataset}
\label{sec:datasets}
We run our experiment and analysis on the TREC DL 2023 passage ranking\footnote{\url{https://microsoft.github.io/msmarco/TREC-Deep-Learning.html}} dataset \cite{craswell2024overview,rahmani2024synthetic}. To the best of our knowledge, this is the only existing test collection dataset that contains human and synthetic queries as well as human expert and LLM judgment annotations. The dataset contains 82 queries ($51$ human/real, $18$ GPT-4-generated, $13$ T5-generated), with judgments on a 4-point scale. The dataset includes 1,830 perfectly relevant (3), 2,259 highly relevant (2), 4,372 related (1), and 13,866 irrelevant judgments.
\section{Human vs.~Synthetic Query Analysis}
\label{sec:analysis}
In this section, we analyse the characteristics of real (or human) vs.~synthetic queries. For analyzing at the query level, two such characteristics are \textit{query length} and \textit{query vocabulary}.

\begin{table}[]
\centering
\caption{The frequent words in real and synthetic queries from TREC DL 2023.}
\begin{tabular}{lrrrrrr}
\toprule
& \multicolumn{3}{c}{Real Queries} & \multicolumn{3}{c}{Synthetic Queries} \\
\cmidrule(lr){2-4} \cmidrule(lr){5-7}
Word & Count & \% & [Rank] & Count & \% & [Rank]\\
\midrule
what & 21 & 7.14\% & [1] & 10 & 3.75\% & [2] \\
is & 14 & 4.76\% & [2] & 6 & 2.25\% & [7] \\
how & 13 & 4.42\% & [3] & 3 & 1.12\% & [12] \\
of & 9 & 3.06\% & [4] & 7 & 2.62\% & [4] \\
in & 8 & 2.72\% & [5] & 7 & 2.62\% & [3] \\
to & 8 & 2.72\% & [6] & 3 & 1.12\% & [13] \\
a & 7 & 2.38\% & [7] & 6 & 2.25\% & [8] \\
the & 7 & 2.38\% & [8] & 15 & 5.62\% & [1] \\
does & 4 & 1.36\% & [9] & 6 & 2.25\% & [6] \\
did & 4 & 1.36\% & [10] & 0 & 0\% & [-] \\
\bottomrule
\end{tabular}
\label{tab:first-word}
\end{table}

Figure \ref{fig:query-length-precentage} shows that real queries tend to have fewer words compared to synthetic queries, suggesting that real user-generated queries are generally more concise. This difference in length may reflect natural tendencies in user behavior, where users aim to minimize typing effort or expect search engines to handle incomplete or less-detailed questions effectively. Table \ref{tab:first-word} provides a comparison of the initial words used in queries. It shows that real queries are significantly more likely to begin with the word `\textit{what}' (7.4\% vs.~3.75\%), and real queries contain a greater number of `what' questions overall. This suggests that real users prefer to ask direct, fact-seeking questions, possibly due to the nature of the underlying task or information need, which synthetic models may not fully replicate. Both real and synthetic query sets, however, display a relatively balanced distribution of common terms, indicating that synthetic queries are capturing some aspects of real queries effectively.

\section{Human vs.~Synthetic Judgment Analysis}
\label{sec:label_analysis}
This section analyzes the difference between the human relevance judgment and the relevance judgment generated by LLMs.

\subsection{Bland Altman Bias Analysis}
\label{sec:bland-altman}
Figure \ref{fig:label_diff} presents a Bland-Altman plot, which is used to visually assess the agreement between the judgments made by the LLM and human evaluators. The plot reveals a bias of approximately $0.28$, indicating that, on average, the LLM tends to assign slightly higher relevance scores compared to human judges. This small negative bias suggests a consistent but minor tendency for the LLM to be more lenient in its scoring (i.e., tends to overestimate). The 95\% Limits of Agreement, which range from approximately $-2.1$ to $1.62$, capture the expected variability in the differences between the LLM and human scores. These limits suggest that in most cases, the difference in scores between the LLM and the human evaluators lies within this range. The relatively wide span of the limits indicates substantial variability, implying that while the LLM's judgments are often close to those of human evaluators, there are instances where they significantly diverge.

The distribution of points is widely spread within these limits; it may suggest that the LLM's scoring behavior varies depending on the query's difficulty or the passage's characteristics. Additionally, the presence of any outliers -- points lying outside the $95\%$ limits -- can highlight cases where the LLM's judgment significantly deviates from human evaluators.

\subsection{Difference in Distributions}

\begin{figure}[t]
    \centering
    \includegraphics[width=\linewidth]{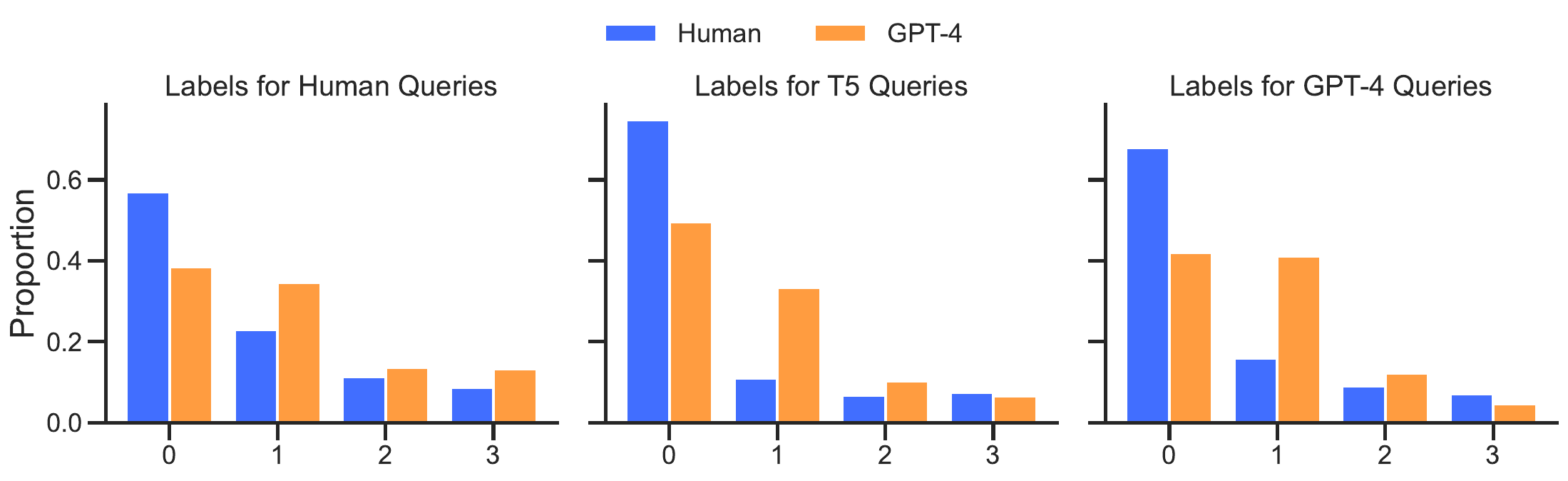}
    \caption{Distribution of relevance labels}
    \label{fig:label_bar}
\end{figure}

Figure \ref{fig:label_bar} shows the distribution of relevance labels assigned by human annotators and GPT-4 across three query types: Human, T5, and GPT-4 queries. Across all query types, human annotators consistently assign a higher proportion of label 0 (irrelevant), indicating a more conservative approach to judging relevance. In contrast, GPT-4 more frequently selects label 1 (relevant), suggesting a tendency to interpret borderline cases as having at least some relevance. This shift is especially evident for T5 queries, where GPT-4 assigns significantly fewer label 0s and more label 1s compared to human annotations.
Higher relevance labels (2 and 3) appear less frequently overall, and the differences between human and GPT-4 judgments are generally smaller and more varied between query types. GPT-4 gives slightly higher for label 2 and slightly lower for label 3. This pattern implies that while GPT-4 is more lenient in identifying weak relevance, it is more hesitant to assign strong relevance scores. The differences across query types and annotator sources underscore systematic variations in how relevance is judged, highlighting potential biases introduced by LLM-based labeling and the influence of query origin on perceived relevance.

The Kullback-Leibler (KL) divergence was also calculated to quantify the difference between the distribution of relevance label distributions generated by GPT-4 compared to human labels. The lower KL divergence indicates closer alignment with human labels, while higher values suggest greater divergence. Overall, the KL divergence results indicate that GPT-4 consistently align  closely with human relevance judgements across all query types ($KL=0.0186$), with similar divergence shown for human and GPT-4 queries ($KL=0.0328$, $KL=0.0335$), but slightly higher for T5 queries ($KL=0.0509$). 

\section{System Ranking Analysis}
\label{sec:system-ranking-analysis}

\begin{figure*}[t]
    \centering
    \subfloat[Synthetic Queries\label{fig:ndcg-syntheticXhuman}]
    {{\includegraphics[scale=0.265]{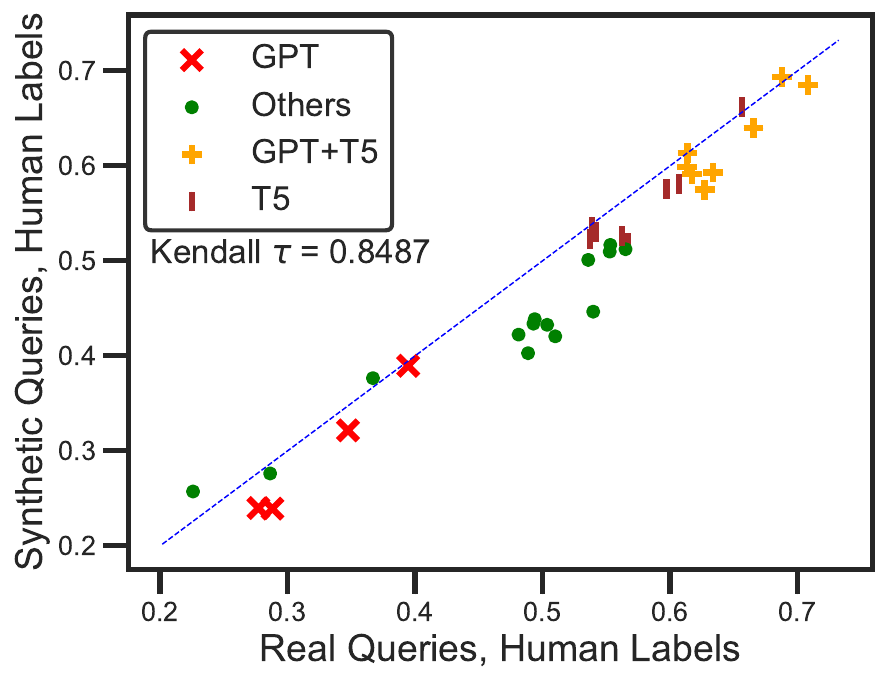}}}%
    \subfloat[GPT-4 Labels\label{fig:ndcg-realXgpt4}]
    {{\includegraphics[scale=0.265]{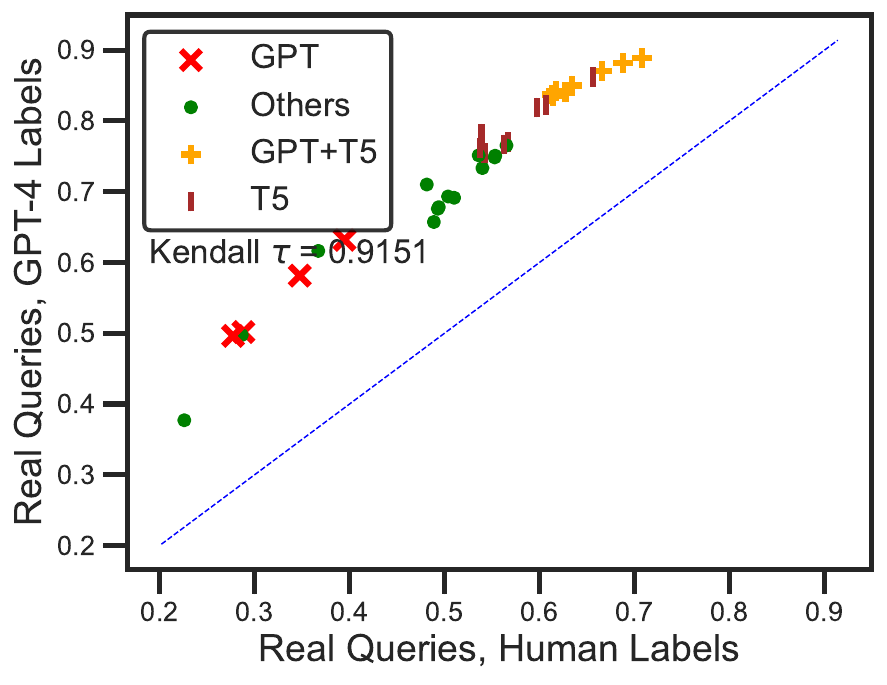}}}%
    \subfloat[GPT-4 Queries\label{fig:ndcg-gpt4Xhuman}]
    {{\includegraphics[scale=0.265]{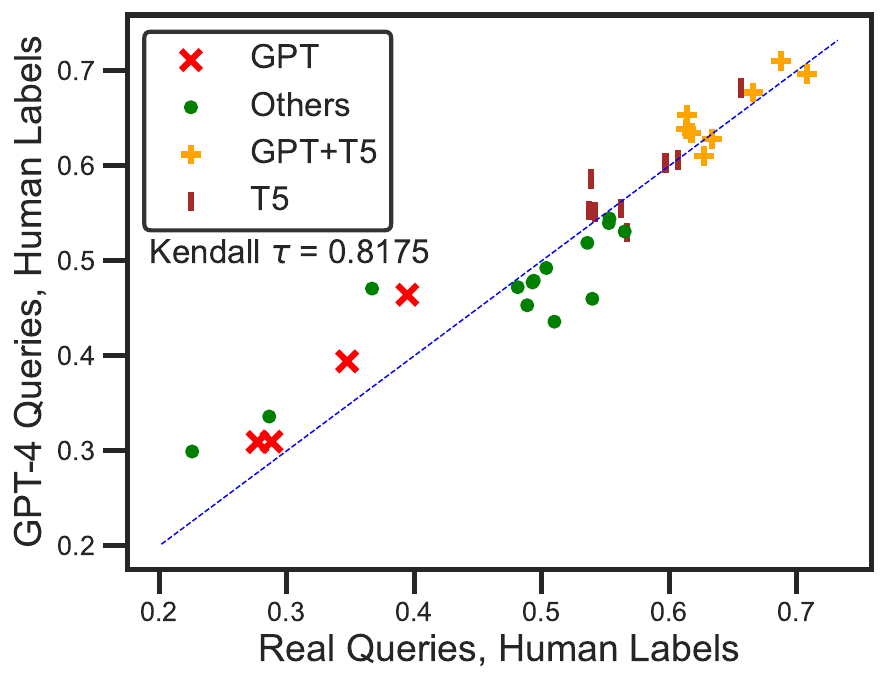}}}%
    \subfloat[T5 Queries\label{fig:ndcg-realXt5}]
    {{\includegraphics[scale=0.265]{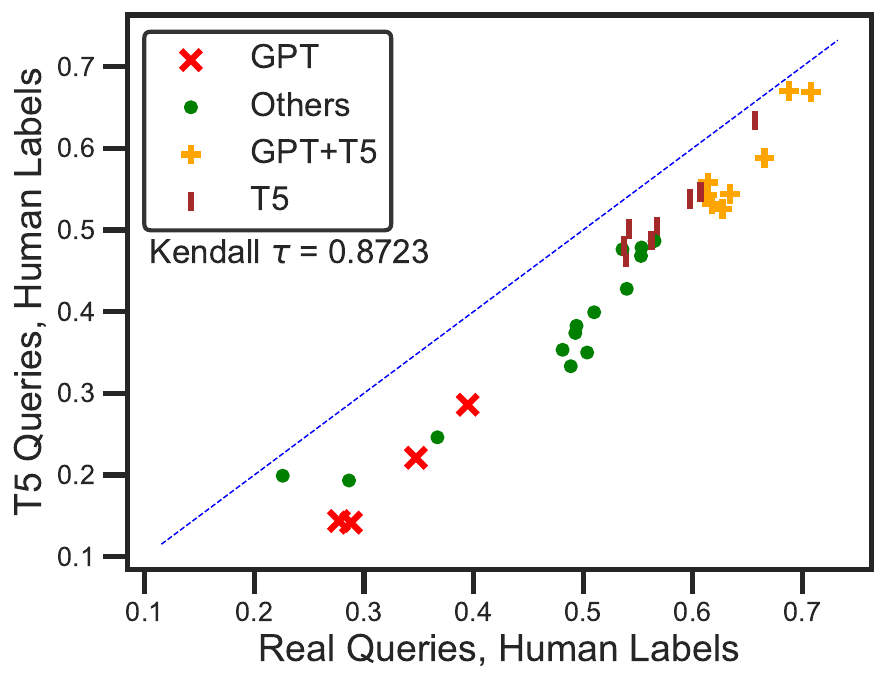}}}%
    \quad
    \subfloat[Synthetic Queries\label{fig:map-syntheticXhuman}]
    {{\includegraphics[scale=0.265]{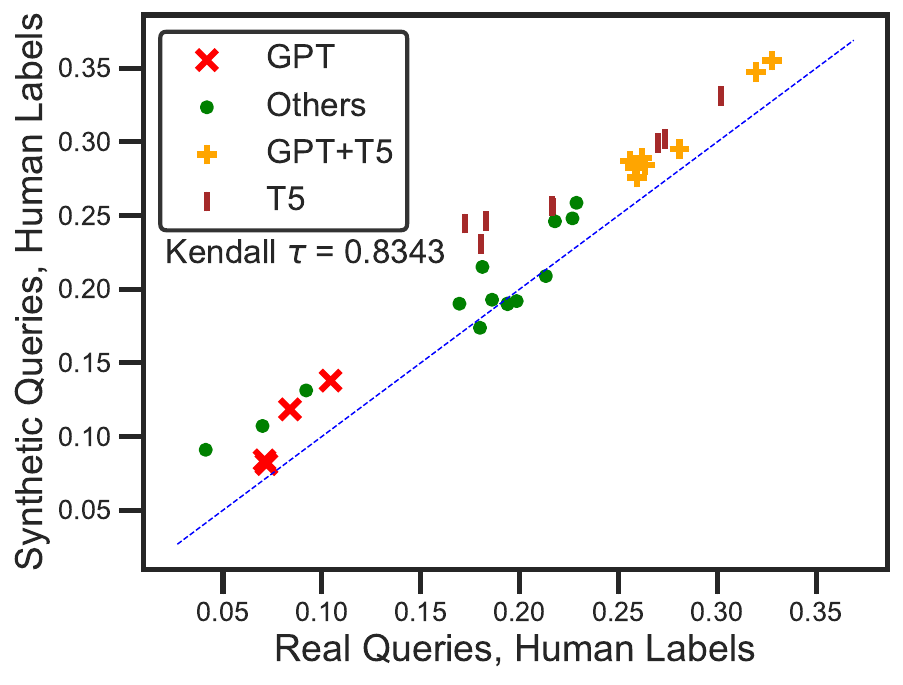}}}%
    \subfloat[GPT-4 Labels\label{fig:map-realXgpt4}]
    {{\includegraphics[scale=0.265]{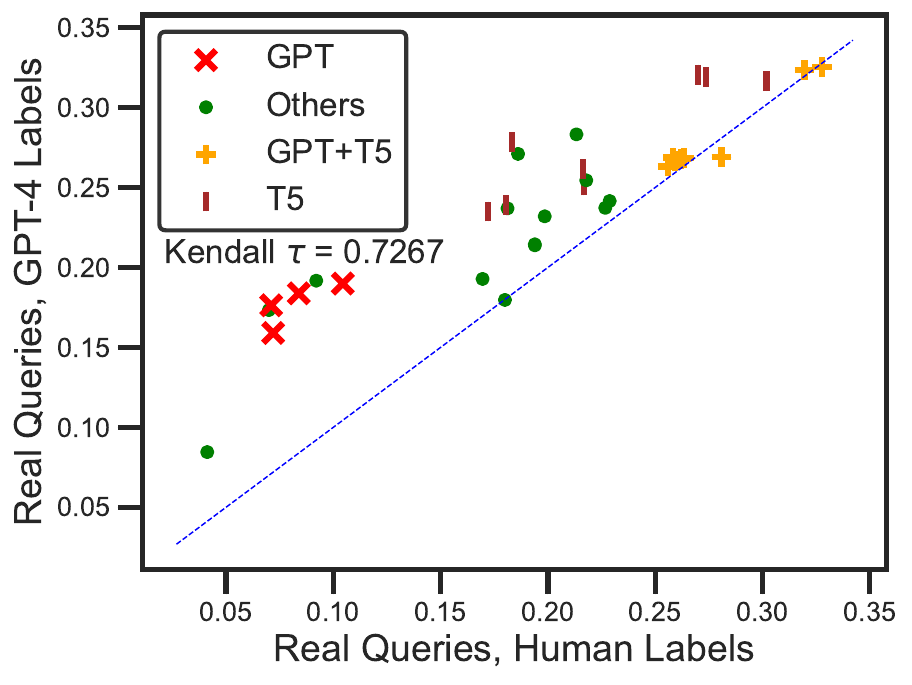}}}%
    \subfloat[GPT-4 Queries\label{fig:map-gpt4Xhuman}]
    {{\includegraphics[scale=0.265]{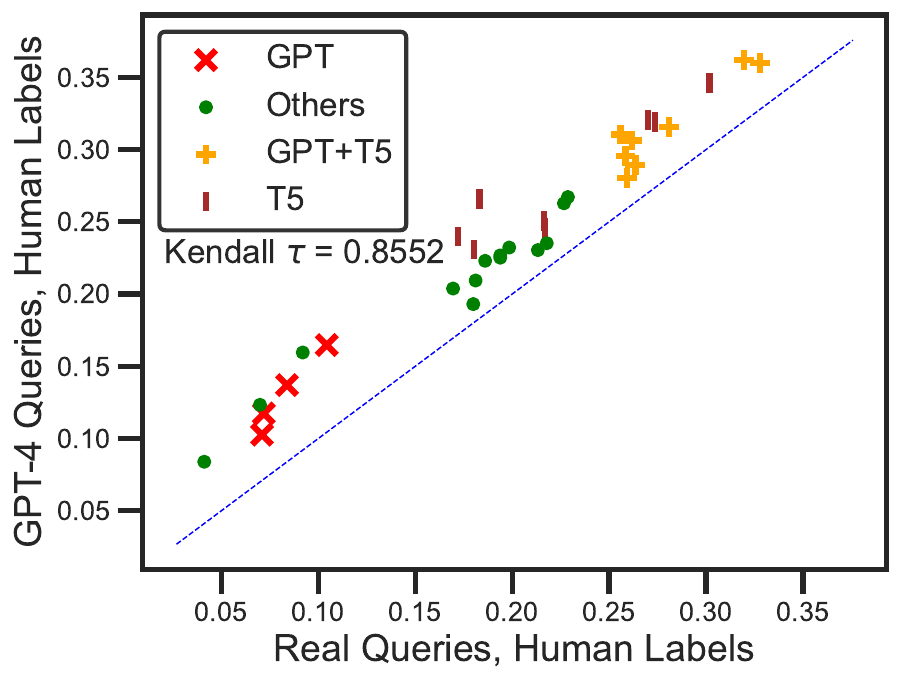}}}%
    \subfloat[T5 Queries\label{fig:map-realXt5}]
    {{\includegraphics[scale=0.265]{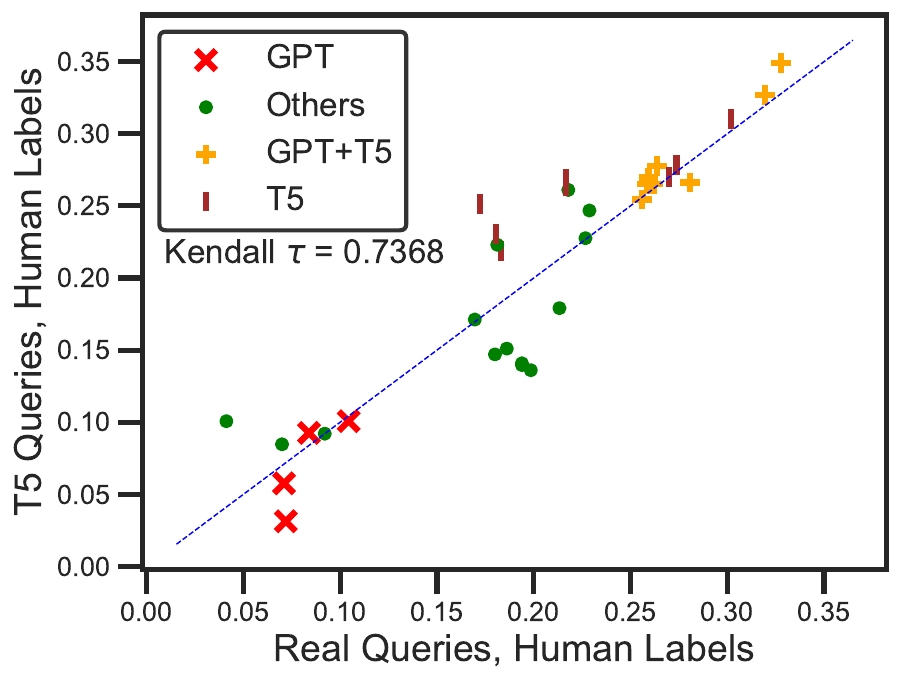}}}%
    \caption{Scatter plots of the effectiveness of TREC Deep Learning Track 2023 runs based on the generated synthetic evaluation test collection. Comparison of various human and synthetic configurations using NDCG@10 (top) and MAP (bottom).}
    \label{fig:system-eval}%
\end{figure*}

In this section, we analyse the bias that a test collection may exhibit towards systems using language models similar to those employed during its construction. Following \cite{rahmani2024synthetic}, we categorised TREC DL 2023 systems by their underlying models using the SynDL metadata \cite{rahmani2024syndl}. This yields four categories: \textbf{GPT}-based, \textbf{T5}-based, \textbf{GPT + T5} (combined), and \textbf{others} (e.g., BM25 or models not using GPT/T5). Figure \ref{fig:system-eval} shows that across all system types, we observe that system performance tends to be higher on synthetic queries compared to real queries, indicating that synthetic test collections, regardless of whether they are constructed using GPT-4 or T5, may be easier and thus overestimate system effectiveness. For example, in Figures \ref{fig:ndcg-realXgpt4} and \ref{fig:map-realXgpt4}, where synthetic labels are compared to real queries under human labels, nearly all system types show a consistent upward shift, suggesting an overoptimistic view of their true performance. This pattern is particularly noticeable in GPT-only runs, which often appear above the diagonal, highlighting their tendency to benefit from synthetic setups.

The results point to a systematic bias in favor of systems that match the model used in test collection generation. For example, GPT-4-based systems are consistently favored when evaluated on GPT-4 queries or GPT-4 labels (see Figures \ref{fig:ndcg-gpt4Xhuman} and \ref{fig:map-gpt4Xhuman}). However, the Kendall's $\tau$ values further show moderate to high correlations, with hybrid setups like GPT+T5 showing better alignment with human-based evaluations. While encouraging, these findings are based on a single test collection, and more extensive experiments are needed to confirm whether these trends hold more generally, especially in different domains or under alternative prompting strategies.
\section{System Evaluation Score Analysis}
\label{sec:system_eval}
We use NDCG and MAP metrics to evaluate various IR systems, comparing traditional approaches to those incorporating large language models (LLMs) in the retriever pipeline. In this section, we analyse the impact of using LLMs in the retrieval process, focusing on how their inclusion affects performance in relation to LLM-generated labels and LLM-generated queries.

\subsection{Linear Mixed Model} 

\subsubsection{Modeling}
To assess differences in metric scores across various properties of retrieval systems, queries, and passages, we fit the data using a linear mixed-effects model. The model examines how the score depends on whether the judgment was made by GPT-4, the system type (GPT-based, T5-based, a combination of both, or traditional), and other system and query features. These include whether the query was generated by GPT-4, query difficulty (QDR), query length (QW), the average document length (DL), and the number of models used in the system pipeline (MN). To capture potential systematic differences in how GPT-4 evaluates outputs compared to human judges, we include interaction terms between the GPT-4 judging indicator and each of the other variables. A random intercept for each retrieval system (i.e., submission runs) accounts for differences and handles multiple observations. The model is estimated separately for two dependent variables, the NDCG and the MAP metrics. The modeling is detailed mathematically below,

{\footnotesize
\begin{displaymath}
\begin{aligned}
Y_{ij} = \; & \beta_0 + \beta_1 \, \text{JudgeGPT4}_{ij} \\
& + \beta_2 \, \text{SysType}_{ij,\text{GPT}} + \beta_3 \, \text{SysType}_{ij,\text{T5}} + \beta_4 \, \text{SysType}_{ij,\text{T5+GPT}} \\
& + \beta_5 \, \text{QDR}_{ij} + \beta_6 \, \text{QW}_{ij} + \beta_7 \, \text{DL}_{ij} + \beta_8 \, \text{isGPT4}_{ij} + \beta_9 \, \text{MN}_{ij} \\
& + \beta_{10} \left(\text{JudgeGPT4}_{ij} \times \text{SysType}_{ij,\text{GPT}} \right) + \beta_{11} \left(\text{JudgeGPT4}_{ij} \times \text{SysType}_{ij,\text{T5}} \right) \\
& + \beta_{12} \left(\text{JudgeGPT4}_{ij} \times \text{SysType}_{ij,\text{T5+GPT}} \right) + \beta_{13} \left(\text{JudgeGPT4}_{ij} \times \text{QDR}_{ij} \right) \\
& + \beta_{14} \left(\text{JudgeGPT4}_{ij} \times \text{QW}_{ij} \right) + \beta_{15} \left(\text{JudgeGPT4}_{ij} \times \text{DL}_{ij} \right) \\
& + \beta_{16} \left(\text{JudgeGPT4}_{ij} \times \text{isGPT4}_{ij} \right) + \beta_{17} \left(\text{JudgeGPT4}_{ij} \times \text{MN}_{ij} \right)
+ \alpha_j + \varepsilon_{ij}
\end{aligned}
\end{displaymath}
}
{\footnotesize
\begin{displaymath}
\alpha_j \sim \mathcal{N}(0, \sigma_\alpha^2), \quad \varepsilon_{ij} \sim \mathcal{N}(0, \sigma^2)
\end{displaymath}
}

where \(i\) indexes observations (i.e., queries) and \(j\) indexes different retrieval runs. The term \(\alpha_j\) is a random intercept capturing variation across runs, assumed normally distributed with variance \(\sigma_\alpha^2\). Residual errors \(\varepsilon_{ij}\) are also normally distributed with variance \(\sigma^2\). \(Y_{ij}\) is the score for observation \(i\). \(\text{JudgeGPT4}_{ij}\) is a binary indicator for whether the judge is GPT-4 (reference: human judge). \(\text{SysType}_{ij,k}\) are dummy variables for system types \(k \in \{\text{GPT}, \text{T5}, \text{T5+GPT}\}\) (reference: traditional). \(\text{QDR}_{ij}\) is query difficulty, \(\text{QW}_{ij}\) is query length by words, \(\text{DL}_{ij}\) is average document length, \(\text{isGPT4}_{ij}\) indicates whether the query was generated by GPT-4, and \(\text{MN}_{ij}\) is the number of models used in the system pipeline. 

\begin{table}[t]
\centering
\caption{
Summary of Linear Mixed Model Coefficient Estimates for NDCG and MAP Scores. The models are based on 5740 observations, evaluating the effects of various predictors. Regression coefficients, standard errors, and p-values are presented. Significant coefficients (p < 0.05) are in \textbf{bold}.}
\begin{adjustbox}{width=\columnwidth}
\begin{tabular}{lcccccc}
\toprule
\multirow{2}{*}{\textbf{Variable}} & \multicolumn{3}{c}{\textbf{MAP Score}} & \multicolumn{3}{c}{\textbf{NDCG Score}} \\
\cmidrule(lr){2-4} \cmidrule(lr){5-7}
 & \textbf{Coef.} & \textbf{Std. Err.} & \textbf{p-value} & \textbf{Coef.} & \textbf{Std. Err.} & \textbf{p-value} \\
\midrule
Intercept & \textbf{0.121} & 0.016 & \textbf{0.000} & \textbf{0.412} & 0.030 & \textbf{0.000} \\
Judged by GPT-4 & \textbf{0.069} & 0.016 & \textbf{0.000} & \textbf{0.207} & 0.021 & \textbf{0.000} \\
System Type: GPT & \textbf{-0.051} & 0.019 & \textbf{0.008} & -0.076 & 0.039 & 0.054 \\
System Type: T5 & -0.001 & 0.023 & 0.964 & 0.034 & 0.047 & 0.466 \\
System Type: T5+GPT & \textbf{0.092} & 0.018 & \textbf{0.000} & \textbf{0.166} & 0.036 & \textbf{0.000} \\
Query Difficulty & \textbf{-0.014} & 0.002 & \textbf{0.000} & \textbf{0.033} & 0.003 & \textbf{0.000} \\
Query Length & 0.001 & 0.001 & 0.376 & -0.004 & 0.002 & \textbf{0.021} \\
Document Length & \textbf{0.000} & 0.000 & \textbf{0.000} & -0.000 & 0.000 & 0.482 \\
GPT4 Query & \textbf{0.030} & 0.010 & \textbf{0.002} & \textbf{0.059} & 0.013 & \textbf{0.000} \\
MN & \textbf{0.016} & 0.004 & \textbf{0.000} & \textbf{0.019} & 0.008 & \textbf{0.016} \\
\midrule
\textbf{Interactions} & & & & & & \\
Judged by GPT-4: SysType (GPT) & \textbf{0.032} & 0.012 & \textbf{0.006} & 0.018 & 0.016 & 0.254 \\
Judged by GPT-4: SysType (T5) & -0.001 & 0.014 & 0.927 & 0.008 & 0.019 & 0.688 \\
Judged by GPT-4: SysType (T5+GPT) & \textbf{-0.027} & 0.011 & \textbf{0.014} & 0.003 & 0.015 & 0.827 \\
Judged by GPT-4: Query Difficulty & 0.001 & 0.003 & 0.653 & \textbf{-0.009} & 0.004 & \textbf{0.049} \\
Judged by GPT-4: Query Length & \textbf{-0.011} & 0.002 & \textbf{0.000} & \textbf{0.006} & 0.003 & \textbf{0.027} \\
Judged by GPT-4: Document Length & \textbf{0.000} & 0.000 & \textbf{0.000} & -0.000 & 0.000 & \textbf{0.000} \\
Judged by GPT-4: GPT4 Query & \textbf{0.026} & 0.013 & \textbf{0.049} & \textbf{-0.040} & 0.018 & \textbf{0.030} \\
Judged by GPT-4: MN & 0.001 & 0.002 & 0.732 & 0.000 & 0.003 & 0.952 \\
\bottomrule
\end{tabular}
\end{adjustbox}
\label{tab:score_model_new}
\end{table}

\subsubsection{Results}
\tabref{tab:score_model_new} presents the model and coefficient summary results. Across both MAP and NDCG scores, systems receive significantly higher relevance scores when judged by GPT-4 compared to human evaluators (MAP: $\beta = 0.069$, $p < 0.001$; NDCG: $\beta = 0.207$, $p < 0.001$). This pattern indicates a general \textit{positive bias} in GPT-4’s assessments relative to human judgments. Moreover, GPT-4 exhibits a small but measurable preference for systems similar to itself. Specifically, GPT-4 assigns slightly higher MAP scores to GPT-4-based systems compared to other system types (interaction $\beta = 0.032$, $p = 0.006$). Conversely, GPT-4 judgments are associated with a slight penalization on MAP scores for hybrid systems combining T5 and GPT components (interaction $\beta = -0.027$, $p = 0.014$).

GPT-4’s evaluations are also sensitive to characteristics of the input queries and documents. When GPT-4 acts as the evaluator, longer queries correlate with a slight decrease in MAP scores (coefficient = --0.011, $p < 0.001$) but a slight increase in NDCG scores (coefficient = 0.006, $p = 0.027$), indicating that GPT-4’s scoring behavior depends on the metric used. Additionally, GPT-4 tends to assign higher MAP scores to queries generated by GPT-4 itself (coefficient = 0.026, $p = 0.049$), but lower NDCG scores for those same queries (coefficient = --0.040, $p = 0.030$). These mixed effects demonstrate that GPT-4’s judgments vary with both query features and evaluation metrics.

\section{Conclusion}
\label{sec:conclusion}
In this paper, we investigated the effect of synthetic data in constructing test collections for evaluation in IR. We analyzed the differences between synthetic and human-generated queries and judgments, showing that LLMs tend to produce longer queries and assign higher relevance scores. Our experiments on system rankings revealed that LLM-based systems are often overestimated when evaluated on synthetic test collections, especially when synthetic labels are used. Furthermore, our Linear Mixed-Effects model shows that GPT-4 systematically gives higher MAP and NDCG scores than human judges and exhibits bias toward systems and queries similar to its own outputs. These findings highlight the limitations of synthetic data for fair IR evaluation and underline the importance of incorporating human oversight when using LLMs for judgment generation.

\begin{acks}
This research is supported by the Engineering and Physical Sciences Research Council [EP/S021566/1] and The Alan Turing Institute's Enrichment scheme.
\end{acks}

\section*{GenAI Usage Disclosure}
The use of ChatGPT was used to help with the writing of this paper. An initial text is given to ChatGPT and asked to paraphrase the given text. Further edits were also applied to the generated text.



\bibliographystyle{ACM-Reference-Format}
\balance
\bibliography{references}


\end{document}